\documentclass[eqsecnum,twocolumn,aps,floatfix]{revtex4}
\usepackage{graphicx}
\begin{document}
\title{Viscous fluid dynamics}
\author{A. K. Chaudhuri}
\email[E-mail:]{akc@veccal.ernet.in}
\affiliation{Variable Energy Cyclotron Centre, 1/AF, Bidhan Nagar, 
Kolkata 700~064, India}

\begin{abstract}
We briefly discuss the phenomenological theory of dissipative fluid. We also present some numerical results for hydrodynamic evolution of QGP fluid with dissipation due to shear viscosity only. Its effect on particle production is also studied. 
\end{abstract}


\date{\today}  

\maketitle

\section{Introduction}
 
A large volume of experimental data from Au+Au collisions at RHIC are successfully analysed in an {\em ideal} fluid dynamic model \cite{QGP3}.   However, experimental data do show deviation from ideal behavior.  The ideal
fluid description works well in almost central Au+Au collisions near  mid-rapidity at top RHIC energy, but gradually breaks
down in more  peripheral collisions, at forward rapidity, or at lower collision energies \cite{Heinz:2004ar}, indicating the
onset of dissipative effects. To describe such deviations from ideal fluid dynamics, quantitatively,  requires the  numerical
implementation of {\em dissipative} relativistic fluid dynamics.
  
Though the theories of dissipative hydrodynamics \cite{Eckart,LL63,IS79} 
has been known for more than 30 years , significant progress towards its numerical implementation  has only been made very
recently \cite{Teaney:2003kp,MR04,CH05,Heinz:2005bw,Chaudhuri:2006jd}.
  In the following, I will briefly review  the phenomenological theory of dissipative hydrodynamics. Some numerical results comparing the ideal and 1st order dissipative hydrodynamics, obtained using the computer code AZHYDRO-KOLKATA \cite{Chaudhuri:2006jd} will be shown.

\section{Phenomenological theory of dissipative hydrodynamics}

In this section, I briefly discuss the phenomenological theory of
dissipative hydrodynamics. More detailed exposition can be found in \cite{IS79}.  

A simple fluid, in an arbitrary state, is fully specified by primary variables: particle current ($N^\mu$), energy-momentum tensor ($T^{\mu\nu}$) and
entropy current ($S^\mu$) and a number of additional (unknown) variables. Primary variables satisfies the conservation
laws;

\begin{eqnarray} 
\partial_\mu N^\mu =&&0,\label{1a}\\
\partial_\mu T^{\mu\nu}=&&0, \label{1b}
\end{eqnarray} 

\noindent and the 2nd law of thermodynamics,

\begin{equation}
\partial_\mu S^\mu  \geq 0.
\end{equation}

In relativistic fluid dynamics, one defines a time-like hydrodynamic 4-velocity,
$u^\mu$ (normalised as $u^2=1$). One also define 
a projector, 
$\Delta^{\mu\nu}=g^{\mu\nu}-u^\mu u^\nu$,
orthogonal to the 4-velocity ($\Delta^{\mu\nu}u_\nu=0$).
In equilibrium, an unique
4-velocity ($u^\mu$) exists such that the particle density ($n$), energy density ($\varepsilon$) and the entropy density ($s$) can be obtained from,

\begin{eqnarray} \label{eq4a}
 N^\mu_{eq}=&& n u_\mu  \\
\label{eq4b}
T^{\mu\nu}_{eq}=&&\varepsilon u^\mu u^\nu -p \Delta^{\mu\nu}\\
\label{eq4c}
S^\mu_{eq}=&&s u_\mu 
\end{eqnarray}

An equilibrium state is assumed to be fully specified by 5-parameters,
$(n,\varepsilon,u^\mu)$ or equivalently by the thermal potential,
$\alpha=\mu/T$ ($\mu$ being the chemical potential) and inverse 4-temperature, $\beta^\mu=u^\mu/T$. Given a equation of state, $s=s(\varepsilon,n)$, pressure $p$ can be obtained from the generalised thermodynamic relation,

\begin{equation} \label{eq6}
S^\mu_{eq}=p\beta^\mu-\alpha N^\mu_{eq} +\beta_\lambda T^{\lambda\mu}_{eq}
\end{equation} 

Using the Gibbs-Duhem relation, 
$d(p\beta^\mu)=N^\mu_{eq} d\alpha -T^{\lambda\mu}_{eq}d\beta_\lambda$, following relations can be established on the equilibrium hyper-surface $\Sigma_{eq}(\alpha,\beta^\mu)$,


\begin{equation} \label{eq7}
dS^\mu_{eq}=-\alpha dN^\mu_{eq}+\beta_\lambda dT^{\lambda\mu}_{eq}
\end{equation}

In a non-equilibrium system, no 4-velocity can be found such that Eqs.\ref{eq4a},\ref{eq4b},\ref{eq4c} remain valid. Tensor decomposition leads to additional terms,  

\begin{eqnarray}\label{eq7abc}
N^\mu=&&N^\mu_{eq}+\delta N^\mu=nu^\mu + V^\mu\\
T^{\mu\nu} =&&T^{\mu\nu}_{eq}+\delta T^{\mu\nu}\\
=&&
[\varepsilon u^\mu u^\nu-p\Delta^{\mu\nu}]+\Pi\Delta^{\mu\nu} + \pi^{\mu\nu} \nonumber\\
&&+(W^\mu u^\nu + W^\nu u^\mu)\\
S^\mu=&&S^\mu_{eq}+\delta S^\mu=su^\mu + \Phi^\mu
\end{eqnarray} 

The new terms describe a net flow of charge $V^\mu=\Delta^{\mu\nu} N_\nu$, heat flow, $W^\mu=(\varepsilon+p)/n V^\mu +q^\mu$ (where $q^\mu$ is the heat flow vector), and entropy flow $\Phi^\mu$. 
$\Pi=-\frac{1}{3}\Delta_{\mu\nu}T^{\mu\nu}-p$ is the bulk viscous pressure
and $\pi^{\mu\nu}= [\frac{1}{2}(\Delta^{\mu\sigma}\Delta^{\nu\tau}+ 
\Delta^{\nu\sigma}\Delta^{\mu\tau}-\frac{1}{3}
\Delta^{\mu\nu}\Delta^{\sigma\tau}]T_{\sigma\tau}$ is the shear stress tensor.
Hydrodynamic 4-velocity can be chosen
to eliminate either $V^\mu$ (the Eckart frame, $u^\mu$ is parallel 
to particle flow) or the heat flow $q^\mu$ (the Landau frame, $u^\mu$ is
parallel to energy flow). In relativistic heavy ion collisions, Landau's frame is more appropriate than the Eckart's frame. Dissipative flows are transverse to $u^\mu$ and additionally, shear stress tensor is traceless. Thus a non-equilibrium state  require 1+3+5=9 additional quantities, the dissipative 
flows $\Pi$, $q^\mu$ (or $V^\mu$) and $\pi^{\mu\nu}$.  
In kinetic theory, $N^\mu$ and $T^{\mu\nu}$ are 1st and 2nd moment of the distribution function. Unless the function is known a-priori, two moments do not furnish enough information to enumerate the microscopic states required to determine $S^\mu$, and
in an arbitrary non-equilibrium state, no relation exists between,
$N^\nu$, $T^{\mu\nu}$ and $S^\mu$. 
{\em Only in a state, close to 
a equilibrium one, such a relation can be established}. Assuming that the equilibrium relation Eq.\ref{eq7} remains valid
in a "near equilibrium state" also, the entropy current can be generalised as,

\begin{equation} \label{eq11}
S^\mu=S^\mu_{eq}+dS^\mu
=p\beta^\mu-\alpha N^\mu +\beta_\lambda T^{\lambda\mu} + Q^\mu
\end{equation}

\noindent where $Q^\mu$ is an undetermined quantity in 2nd order in deviations, $\delta N^\mu=N^\mu-N^\mu_{eq}$ and $\delta T^{\mu\nu}=T^{\mu\nu}-T^{\mu\nu}_{eq}$.  Detail form of $Q^\mu$ is constrained by the 2nd law $\partial_\mu S^\mu \geq 0$.
With the help of conservation laws and Gibbs-Duhem relation,
entropy production rate can be written as,

\begin{eqnarray} \label{12}
\partial_\mu S^\mu=-\delta N^\mu \partial_\mu \alpha
+\delta T^{\mu\nu} \partial_\mu \beta_\nu + \partial_\mu Q^\mu
\end{eqnarray}

Choice of $Q^\mu$ leads to 1st order or 2nd order theories of dissipative hydrodynamics.
In 1st order theories $Q^\mu=0$, entropy current contains terms upto 1st order in deviations,
$\delta N^\mu$ and $\delta T^{\mu\nu}$. Entropy production rate can be written as,

\begin{equation}\label{eq213}
T\partial_\mu S^\mu
=\Pi X -q^\mu X_\mu + \pi^{\mu\nu} X_{\mu\nu}  
\end{equation}

\noindent where, $X=-\nabla.u$; $X^\mu=\frac{\nabla^\mu}{T}-u^\nu \partial_\nu u^\mu$ 
and 
$X^{\mu\nu}=\nabla^{<\mu} u^{\nu>}$.
   
The 2nd law, $\partial_\mu S^\mu \geq 0$ can be satisfied by postulating a linear relation between the dissipative flows and and thermodynamic forces,

\begin{eqnarray}
\label{15a}
\Pi=&&-\zeta \theta,\\
\label{15b}
q^\mu=&&-\lambda \frac{nT^2}{\varepsilon+p}\nabla^\mu(\mu/T),\\ 
\label{15c}
\pi^{\mu\nu}=&&2\eta \nabla^{<\mu}u^{\nu>}
\end{eqnarray}

\noindent where $\zeta$, $\lambda$ and $\eta$ are the positive transport coefficients, bulk viscosity, heat conductivity and shear viscosity. 

In 1st order theories, causality is violated. If, in a given fluid cell, at a certain time, thermodynamic forces vanish, corresponding dissipative fluxes also vanish instantly. This is corrected in   2nd order theories \cite{IS79} where entropy current contain terms upto 2nd order in the deviations, $Q^\mu \neq 0$. 
The most general $Q^\mu$ containing terms upto 2nd order in deviations can be written as,

\begin{equation}
Q^\mu=-(\beta_0\Pi^2-\beta_1 q^\nu q_\nu + \beta_2\pi_{\nu\lambda}\pi^{\nu\lambda})
\frac{u^\mu}{2T} -\frac{\alpha_0\Pi q^\mu}{T} +\frac{\alpha_1 \pi^{\mu\nu}q_\nu}{T}
\end{equation}

As before, one can cast the entropy production rate ($T\partial_\mu S^\mu$) in the form of Eq.\ref{eq213}. 
Neglecting the terms involving dissipative flows with gradients of equilibrium thermodynamic quantities (both are assumed to be small) and demanding that a linear
relation exists between the dissipative flows and thermodynamic
forces,  following {\em relaxation} equations for the dissipative flows can be obtained,  

\begin{eqnarray}
\Pi=&&-\zeta (\theta +\beta_0 D\Pi)\\
q^\mu=&&-\lambda \left[ \frac{nT^2}{\varepsilon+p}\nabla^\mu(\frac{\mu}{T}) 
-\beta_1 Dq^\mu \right]\\
\pi^{\mu\nu}=&&2\eta \left[\nabla^{<\mu}u^{\nu>} -\beta_2 D\pi_{\mu\nu} \right],
\end{eqnarray}

\noindent where  $D=u^\mu \partial_\mu$is the convective time derivative. Unlike in the 1st order theories, in 2nd order theories,
 dynamical equations controls the dissipative flows. 
 Even if thermodynamic forces vanish, dissipative flows donot vanish instantly.

\vspace{-.4cm}
\begin{figure}[ht]
  \hfill  
\includegraphics[bb=14 13 581 829,width=0.89\linewidth,clip]{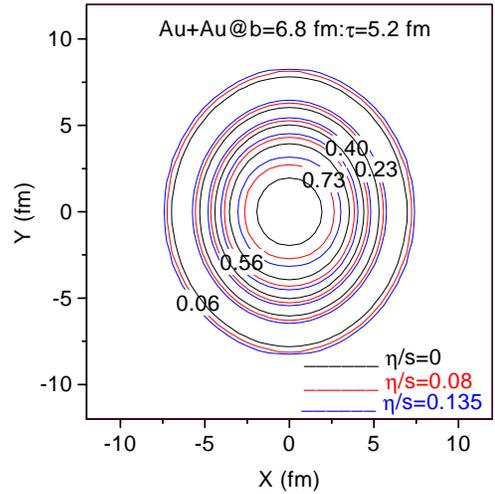}
    \vspace{-3.5cm}
      \caption{Contour plots of energy density at  (proper) time $\tau$=5.2 fm.}
      \label{F1}
  \end{figure}

Before we proceed further, it may be mentioned that the parameters,
$\alpha$ and $\beta_\lambda$ are not connected to the actual state
($N^\mu,T^{\mu\nu}$). The pressure $p$ in Eq.\ref{eq11} is also not
the "actual" thermodynamics pressure, i.e. not the work done in
an isentropic expansion. Chemical potential $\alpha$ and 4-inverse temperature $\beta_\lambda$ has meaning only for the equilibrium state. Their meaning need not be extended to non-equilibrium states also. However, it is possible to define a fictitious "local equilibrium" state, point by point, such that pressure $p$ in
Eq.\ref{eq11} can be identified with the thermodynamic pressure,  
at least upto  1st order. In 2nd order in deviations, such an identification is not possible. 

 \vspace{-.4cm}
\begin{figure}[ht]
  \hfill  
\includegraphics[bb=14 13 581 829,width=0.89\linewidth,clip]{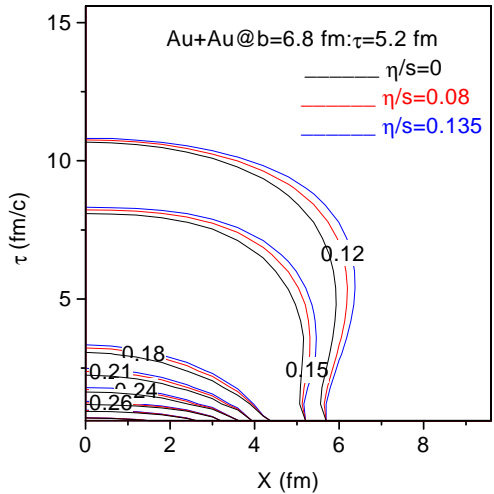}
    \vspace{-3.5cm}
      \caption{Contour plots of temperature at  y=0 in $x-\tau$ plane.}
      \label{F3}
  \end{figure}

\section{Viscous hydrodynamics for QGP in 2+1 dimensions}  

Numerically, causal dissipative hydrodynamics is a challenging problem. One needs to solve simultaneously 14 partial differential 
equations(5 conservation equations and 9 additional equations for the dissipative flows:, $\Pi$, $q^\mu$ and $\pi^{\mu\nu}$
 ). Recently, at the Cyclotron Centre, Kolkata, we have developed a code for solving  causal dissipative hydrodynamics with dissipation due to shear viscosity only, in 1+1 dimension \cite{CH05} (assuming boost-invariance and cylindrical symmetry) and presently extending the code in 2+1 dimensions (with boost-invariance only).  We have completed the coding for the 1st order theory. Here, we present some numerical results.

Assuming boost-invariance, we have solved 1st order viscous hydrodynamics for initial state QGP in
2+1 dimension, in $(\tau,x,y,\eta_s)$ coordinate.  
We restrict ourselves to central rapidity region,
where the QGP fluid is essentially baryon free and to keep the calculations simple, we consider the most
important dissipative term, the shear viscosity
and neglect the other dissipative terms, e.g. heat conduction, bulk viscosity. For viscosity, we have used two values, the ADS/CFT motivated value $\eta/s$=0.08, and the perturbative estimate $\eta/s$=0.135 .
Details of the equations solved can be found in \cite{Chaudhuri:2006jd}.
We just mention that we have used the equation of state,
EOS-Q, developed in ref.\cite{QGP3}, with 1st order phase transition with critical temperature $T_c$=164 MeV.  
As mentioned earlier, 1st order dissipative hydrodynamics 
violate causality and can lead to unphysical effects like early reheating. But for QGP fluid, which is close to an ideal fluid,
such effects can be minimised with appropriate initial conditions,
and we did not find any evidence of early reheating.

\subsection{Evolution of the viscous fluid}
 
In the following we will show the results obtain in Au+Au  collision at impact parameter $b=$ 6.8 fm, which approximately
corresponds
to 16-24\% centrality Au+Au collisions.  With the same initial conditions,
we have solved the energy-momentum conservation equations for
ideal fluid and viscous fluid.  
 In Fig.\ref{F1}, we have shown
the constant energy density contour plot in x-y plane, after an evolution of 
5 fm.
The black lines are for ideal fluid evolution. The  red and blue lines are for
viscous fluid with  $\eta/s$=0.08 and 0.135 respectively. 
   In Fig.\ref{F3},   contour plot of temperature in x-$\tau$ plane is shown. It is evident that compared to ideal fluid, viscous fluid cools slowly.
  Transverse expansion
is also enhanced in a viscous fluid. It is not shown, but  fluid velocity grow  faster in viscous flow.

\begin{figure}[ht]
  \hfill  
\includegraphics[bb=14 13 581 829,width=0.89\linewidth,clip]{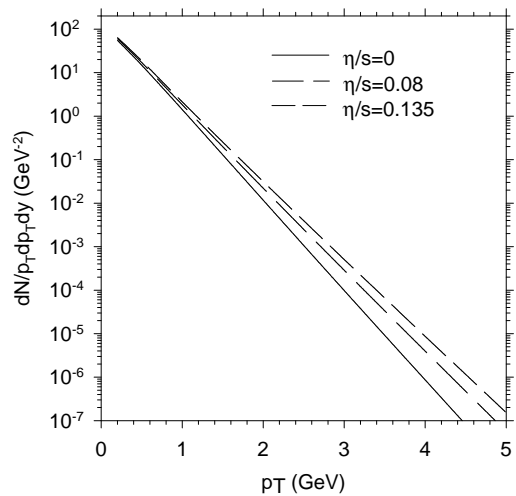}
    \vspace{-4.0cm}
      \caption{$p_T$ spectra of $\pi^-$ for ideal and viscous evolution.}
      \label{F7}
  \end{figure}

\subsection{Particle spectra}

Viscosity
influences the particle production by (i) extending the freeze-out surface and (ii) by introducing
a correction to the
equilibrium distribution function (see \cite{Chaudhuri:2006jd}).
In Fig.\ref{F7}, transverse momentum distribution of 
$\pi^-$ from ideal and viscous hydrodynamics are compared. Freeze-out temperature is $T_F$=0.158 GeV.   Pion production is increased in viscous dynamics.   We also note that effect of viscosity
is more prominent at large $p_T$ than at low $p_T$. $p_T$ spectra of pions are flattened with viscosity. 

\begin{figure}[t]
  \hfill  
\includegraphics[bb=14 13 581 829,width=0.89\linewidth,clip]{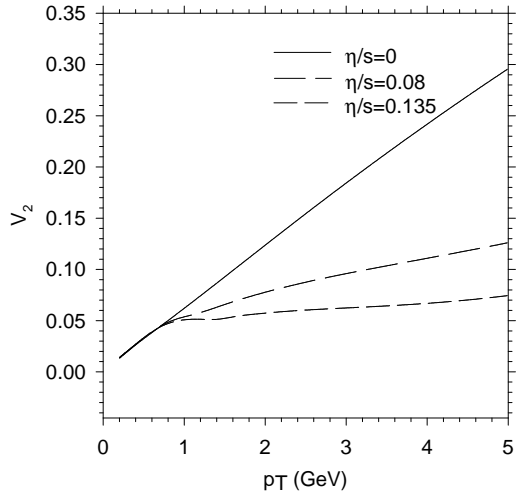}
    \vspace{-4.0cm}
      \caption{$p_T$ dependence of elliptic flow for $\pi^-$.}
      \label{F9}
  \end{figure}

Effect of viscosity is also prominent on elliptic flow (Fig.\ref{F9}). In ideal dynamics,
elliptic flow continues to increase with $p_T$.
 In viscous dynamics, on the otherhand, elliptic flow tends to saturate.
The result is very encouraging, as experimentally also elliptic flow
tends to saturate at large $p_T$.
 
\begin{figure}[t]
  \hfill  
\includegraphics[bb=14 13 581 829,width=0.89\linewidth,clip]{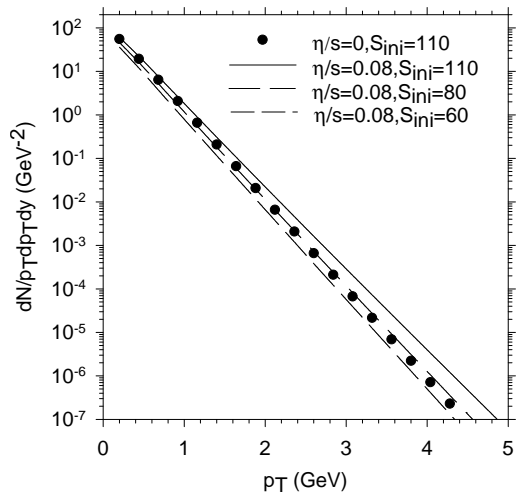}
    \vspace{-4.0cm}
      \caption{$p_T$ spectra of $\pi^-$ from ideal fluid with initial entropy density
110 $fm^{-3}$ is compared with viscous dynamics with different initial entropy density. }
      \label{F10}
  \end{figure}

Lastly, in Fig.\ref{F10}, we have shown a comparison of $p_T$ spectra obtained in ideal hydrodynamics with initial entropy density $S_{ini}$=110 $fm^{-3}$ with $p_T$ spectra obtained
in viscous hydrodynamics obtained with initial entropy density, 60,80 and 110. $p_T$ spectra from viscous fluid initialised with 
$S_{ini}$=60-80 compare well with the spectra from ideal fluid,
initialised at higher entropy density. To produce the same spectra,
viscous fluid require 20-30\% less initial temperature.
 
\section{Summary and conclusions}

We have briefly reviewed the phenomenological theory of dissipative hydrodynamics and presented some numerical results 
from 1st order dissipative hydrodynamic of QGP fluid with only shear viscosity. 1st order theories suffer from the problem of causality, signal can 
travel faster than light.  Unphysical effects like
reheating of the fluid, early in the evolution, can occur.  
In this model study, we have 
considered two values of viscosity, the ADS/CFT motivated value,
$\eta/s \approx$0.08 and perturbatively estimated viscosity, $\eta/s \approx$0.135. We did not find any indication of unphysical reheating.
Explicit simulation of ideal and viscous fluids confirms that energy density of a 
viscous fluid, evolve slowly than its ideal counterpart. The fluid velocities
on the other hand evolve faster in viscous dynamics than in ideal 
dynamics. Transverse expansion is also more in viscous dynamics.

We have also studied the effect of viscosity on particle production.
Viscosity generates entropy leading to enhanced particle production.
Particle production is  increased due to (i) extended freeze-out surface and (ii) non-equilibrium
correction to equilibrium distribution function.  
  With ADS/CFT (perturbative) estimate of viscosity, at $p_T$=3 GeV,
pion production is increased by a factor 3 (5) . Increase is even more at large $p_T$. While viscosity enhances particle
production,
it reduces the elliptic flow. At $p_T$=3 GeV, for ADS/CFT(perturbative) estimate of viscosity,
elliptic flow is reduced by a factor of 2(3). We also find that at large $p_T$
elliptic flow tends to saturate.

To conclude, present study shows viscosity, even if small, can be
very important in analysis of RHIC Au+Au collisions. Currently accepted
initial temperature of hot dense matter produced in RHIC Au+Au collisions,
obtained from ideal fluid analysis
can be changed by 20\% or more with dissipative dynamics.



\begin{thebibliography}{99}
\bibitem{QGP3}
P.~F. Kolb and U. Heinz,
in {\it Quark-Gluon Plasma 3}, edited by R.~C. Hwa and 
X.-N. Wang (World Scientific, Singapore, 2004), p.~634.

\bibitem{Heinz:2004ar}
U.~Heinz,
J.\ Phys.\ G {\bf 31}, S717 (2005).

\bibitem{Eckart}
C.~Eckart, Phys. Rev. {\bf 58}, 919 (1940).

\bibitem{LL63}
L.~D.~Landau and E.~M.~Lifshitz, {\it Fluid Mechanics}, Sect. 127,
Pergamon, Oxford, 1963.

\bibitem{IS79}
W. Israel, Ann. Phys. (N.Y.) {\bf 100}, 310 (1976);
W.~Israel and J.~M.~Stewart, Ann. Phys. (N.Y.) {\bf 118}, 349 (1979).



\bibitem{Teaney:2003kp}
  Phys.\ Rev.\ C {\bf 68}, 034913 (2003)
  [arXiv:nucl-th/0301099].

\bibitem{MR04}
  A.~Muronga and D.~H.~Rischke,
  nucl-th/0407114\,(v2).

\bibitem{CH05}
  A.~K.~Chaudhuri and U.~Heinz,
  nucl-th/0504022.

\bibitem{Heinz:2005bw}
  U.~W.~Heinz, H.~Song and A.~K.~Chaudhuri,
  Phys.\ Rev.\ C {\bf 73}, 034904 (2006)
  [arXiv:nucl-th/0510014].
\bibitem{Chaudhuri:2006jd}
  A.~K.~Chaudhuri,
  Phys.\ Rev.\ C {\bf 74}, 044904 (2006)
  [arXiv:nucl-th/0604014].
\end{thebibliography}
\end{document}